\documentclass{article}

\usepackage{PRIMEarxiv}

\usepackage[utf8]{inputenc} 
\usepackage[T1]{fontenc}    
\usepackage{hyperref}       
\usepackage{url}            
\usepackage{booktabs}       
\usepackage{amsfonts}       
\usepackage{nicefrac}       
\usepackage{microtype}      
\usepackage{lipsum}
\usepackage{fancyhdr}       
\usepackage{graphicx}       
\graphicspath{{media/}}     

\usepackage{amssymb}
\usepackage{amsmath}
\usepackage{amsthm}
\usepackage{mathtools}

\newtheorem{definition}{\bf Definition}[section]
\newtheorem{proposition}{\bf Definition}[section]

\title{Quantum-like states on complex synchronized networks
}

\author{
  Gregory D. Scholes \\
  Department of Chemistry \\
  Princeton University \\
  Princeton, New Jersey, 08540 U.S.A.\\
  \texttt{gscholes@princeton.edu} \\
}

\begin{document}
\maketitle

\begin{abstract}
Recent work has exposed the idea that interesting quantum-like probability laws, including interference effects, can be manifest in classical systems. Here we propose a model for quantum-like (QL) states and QL bits. We suggest a way that huge, complex systems can host robust states that can process information in a QL fashion. Axioms that such states should satisfy are proposed. Specifically, it is shown that building blocks suited for QL states are networks, possibly very complex, that we defined based on $k$-regular random graphs. These networks can dynamically encode a lot of information that is distilled into the emergent states we can use for QL like processing. Although the emergent states are classical, they have properties analogous to quantum states. Concrete examples of how QL functions are possible are given. The possibility of a `QL advantage' for computing-type operations and the potential relevance for new kinds of function in the brain are discussed and left as open questions.
\end{abstract}


\section{Introduction}
Quantum states characterize the microscopic world, where they convey the dual wave-particle basis for quantum theory. Of special interest is the way probability laws for quantum states allow for interference effects and other, more exotic, correlations evidenced in the effect and outcome of measurements. We thus find fascinating non-classical phenomena that can be exploited in quantum information science \cite{NielsenChuang}. Here, a goal is to leverage quantum correlations at the molecular scale for function at the human scale. A well known challenge, however, is that quantum states, at least the most interesting ones, tend to be fragile---easily destroyed by environmental noise and fluctuations (the process of decoherence\cite{Zurek2003}). Researchers therefore focus on stabilizing simple states using techniques such as very low temperature. Here we explore a way to produce robust states, comprising highly complex systems, that exhibit "quantum like" properties. 

Let's start by defining what we mean, in the present work, by a quantum-like (QL) state. An important requirement is that these states allow amplitude level interferences that are explained by a measurement formulation like that required by the quantum theory. This is a fairly broad criterion that can be fulfilled by either: (a) states in vector spaces, where we think of amplitude (vector) additions followed by measurement of probability as the amplitude squared; (b) or we can work only with probabilities, but formulate a vector-like model for measurement sequences (this is the essence of the V\"{a}xj\"{o} model). In the present work we would like to take the QL state framework further, and show how eigenstates of classical systems can be combined (through interactions between the systems serving as QL-bits) to produce states in a state space that represents the tensor product of vector spaces or each QL-bit. Since we are dealing with inner product spaces here, the vector spaces are Hilbert spaces.

Why would complex quantum-like, systems be interesting? There are two major motivations for this work. First, by exhibiting a concrete example of robust quantum-like states we can suggest ways to design circuits with quantum like functions. By considering such circuits, we can conceive of a experiments designed to seek quantum-like correlations in a range of different systems. Moreover, this platform could suggest a way to perform QL computing using simplified hardware that is a step towards true quantum computing. Second, we establish a hypothesis for what quantum-like states could `look like' in settings as complex as biology. The key point we will establish is the feasibility of these states. 

The possibility that some kinds of quantum effects provide function in biological systems has intrigued researchers for over a century. Yet it is easy to argue against such effects because the kinds of quantum states that we are familiar with are fragile, and would rapidly decohere in a noisy environment. This point is well recognized, yet, as enunciated by Penrose\cite{shadows}, calls for discovery of new kinds of quantum systems that are not only robust in a using environment, but "retain their manifest quantum nature at a much larger scale". Even if such states are not present, or functional,  in living systems, elucidating the principles that enable them would open up new ways of engaging quantum (like) correlations for technology. 

In quantum states, classical correlations are prominent\cite{ScholesQIS}. For example, in the entangled state $|\Psi_- \rangle = \frac{1}{\sqrt{2}} \Bigl[ |0\rangle_A |1\rangle_B - |1\rangle_A |0\rangle_B \Bigr] $ it is obvious that if a measurement of qubit A indicates it is in state $|0\rangle_A$, the qubit B (measured in the same basis) must be in the state $|1\rangle_B$. In addition, we note that interference phenomena---a key contributor to quantum function---is not confined to quantum systems. Thus there is reason to believe that classical systems possessing the right kinds of correlations and with the potential to exhibit interference effects  might replicate many (obviously not all) kinds of quantum phenomena. It is thus proposed that it may not always be necessary to produce truly quantum devices that exploit quantum correlations. It might often be sufficient to develop a quantum-like mimic constructed from classical components. Not only does this open new avenues for engineering systems that can capture some of the properties of quantum systems, but it provides a framework for thinking about quantum-like function in nature. The essential advantage would be that more sophisticated probability laws can be leveraged.

Here we combine ideas from graph theory and operator theory to demonstrate how complex networks can serve as a basis for generating complex. quantum-like states and producing functions using those states. We formulate axioms that guide the construction of the states and demonstrate examples of such states. Importantly, we describe how these states are resilient in noisy environments. In the discussion, we speculate about functions unique to these states and how they provide a basis for testable hypotheses for exotic phenomena, including how quantum-like processes might augment classical processes in neural networks like the brain. Some of this might sound for-fetched, but keep in mind that the central finding of this work is to define and analyze the quantum-like states themselves. From this foundation, perhaps we can begin asking questions that could not previously be well posed. 

One inspiration for the work was to define physical realizations of QL states that can show properties described by QL probabilities, as proposed by Khrennikov in the framework of a contextual probabilistic model\cite{Khrennikov1}, termed the V\"{a}xj\"{o} model. The focus of the V\"{a}xj\"{o} model is on non-Kolmogorov probabilities---that is, models where probability can be thought of as a kind of vector quantity. This allows probability amplitude to interfere, like in quantum theory. However, there is no requirement of an underlying microscopic quantum state, so the quantum-like model can be used to analyze problems as far afield as financial markets, psychology, or even the mind. Our goal in this paper is to propose a concrete model for QL states that are complex enough to accommodate the idea of `context' in measurement theory. 

The main focus of the paper is to formulate examples of QL-states in a vector space, closely analogous to quantum states, that can be realized in classical setups. To do that we notice that emergent states of very complex systems are incredibly robust and isolated in the spectrum, so they can potentially serve as building blocks for QL states\cite{ScholesEntropy}. An emergent phenomenon is manifest when numerous small interactions add coherently to produce a stunning collective effect that appears above some threshold of the number of coherently interacting systems. In quantum systems, an emergent state is signalled by a single eigenvalue that splits away from the many other eigenvalues in the spectrum, opening up an energy gap\cite{Fano1992}. The questions that we solve in the present work is how to design networks that produce stable superposition states of two basis states, and how to combine those networks to produce a larger state space that can be mapped onto a state space comprising linear combinations of product states.

Emergent states are characteristic of synchronized complex networks of phase oscillators. There are many examples of synchronization in a wide variety of systems\cite{StrogatzBook, Strogatz2000, Acebron2005, sync1, sync2, sync3, sync4, sync5, sync6}. A few illustrative examples of synchronization include power networks\cite{sync7}, bacterial networks interacting via signalling\cite{sync8}, human networks of violin players\cite{sync9}, and mechanisms underpinning memory processes in the brain\cite{memory}. We will discuss synchronization in neural networks in more detail later in the paper. All these kinds of systems attain a stable synchronized steady state by feedback loops. 

Synchronization is usually analyzed with the Kuramoto Model\cite{Strogatz2000}; a system of coupled differential equations that predict how initial conditions and feedback acting in a set of phase oscillators allow synchronization to emerge. Here we assume that the initial conditions (phase offsets) allow the system to be synchronizable. We have studied the connections between network synchronization and homomorphic quantum eigenstates in other work\cite{Scholes2020}. This approach allows us to focus on the eigenstates rather than dynamics. In particular, we are interested in an ordered eigenstate whose eigenvalue is well separated from the eigenvalues of the many other random states of a system. We construct the basis for these emergent states in terms of networks (or graphs). 

\begin{figure}[!h]
	\centering\includegraphics[width=2.5in]{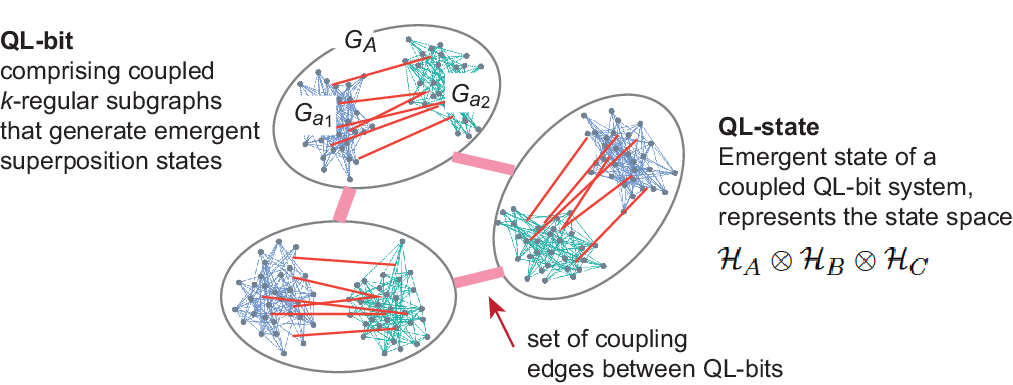}
	\caption{ Example of the way three QL-bits couple to produce any state in the tensor product space. The QL-bits are comprised of graphs, labeled here for later reference. Connections (edges) between the QL-bits are shown schematically. The emergent state of the network produces a state that can be mapped to the a state in the tensor product of Hilbert spaces of each QL-bit, as indicated. In reality, there should be many edges between each pair of subgraphs.  }
	\label{fig_1}
\end{figure}

The main outcome of this paper is to combine these ideas---QL states and emergent states of networks---to propose a network structure that serves as a building block for the full tensor product representation of QL states. That is, we show how to mimic linear combinations of product states. A schematic of the construction that will be developed in the paper is shown in Fig 1. We then discuss how those states might be exploited for function in ways that have not been considered previously for typical network structures. 

Detailed reviews of quantum information science can be found in various books and reviews \cite{NielsenChuang, Ballentine, Peres, Barnett, Horodecki2009, Vedral2002}.  In recent review-type papers\cite{WeijunQIS1, ScholesQIS} we  describe the idea of entanglement and qubits and summarize the conceptual ideas behind quantum information science for a broader readership.

\section{Graphs and Emergent States}
A network can be represented as a graph. A graph $G(n,m)$, that we will often write simply as $G$, comprises $n$ vertices and a set of $m$ edges that connect pairs of vertices. The size of a graph or subgraph, that is, the number of vertices is written $|G|$. In the present work, the edges are assigned semi-randomly---that is, there is an overarching `rule' that guides how the edges are placed randomly. In the present work, we only consider that `rule' to be that we construct $k$-regular random graphs (we explain these graphs below). In some cases, we also delete some fraction of the edges randomly after building the graph\cite{ScholesEntropy}. Rather than assigning a specific number of edges, sometimes we assign  edges between vertices with probability $p$, so the graph is formally $G(n,p)$. Here we discuss only graphs with no loops and no multiple edges. For background see \cite{Diestel,Janson2000, Bollobas2001}.

The adjacency matrix of a graph $A$ is the $n \times n$ matrix containing entries $a_{ij} = a_{ji} = 1$, with $i, j \in \{ 1, 2, \dots n\}$, when an \textit{undirected} edge joins vertices $i$ and $j$. The adjacency matrix of a \textit{directed} graph contains entries $a_{ij} = 1$  when an edge connects from vertex $i$ to vertex $j$. We consider also \textit{signed} graphs, where the edge entry in the adjacency matrix is designated from $\{ 0, 1, -1 \}$.

Undirected graphs can be thought of as representing coupling between entities located at the vertices---a kind of coupling map of the system. For instance, below we will discuss networks of interacting oscillating dipoles, so the interaction is the dipole-dipole coupling. That coupling can be negative, indicating that the coupled oscillators are locked out-of-phase, then we use the signed graph. Directed graphs indicate a flow, timing, or causal coupling from one vertex to another. That is often found in physical systems\cite{Lodi2021}, including transportation networks, aspects of social networks, downstream influence (common in biology), and communication.

The spectrum of a graph $G$ is defined as the spectrum (i.e. eigenvalues) of its adjacency matrix $A$. Here we use the spectrum to detect and characterize emergent states by their eigenvalues. The spectrum can tell us a lot about the graph, and \textit{vice versa}, see for example \cite{graphevals}. The graph can also be used as a template to construct an ensemble of random matrices, whose spectrum is analyzed\cite{Scholes2020}.

In earlier work we noticed that certain graph types have a spectrum where the lowest (or highest, depending on the signs of the edges) eigenvalue (or few eigenvalues) are sufficiently separated from the rest of the spectrum so that those states are resilient to decoherence\cite{Scholes2020}. Indeed, it is known that there are families of graphs with exceptional properties of this kind---expander graphs\cite{Sarnak2004, expandersguide, Lubotzky, Expanders, Expanders2, Alon1986}. In the present work, expander graphs of a particular type will underlie our prototypical networks. Expander graphs involve high connectivity throughout the graph, and are therefore optimal structures for communication or random walks.

\begin{definition}
	(Expander graph) Let $G$ be the $n$-vertex graph with vertex set $V$ and take $0 < \epsilon \in \mathbb{R}$. $G$  is an $\epsilon$-expander if for every vertex subset $Y$ of $V$ with $|Y| \le \tfrac{1}{2}|V|$ we have
	\begin{equation}
		|\partial Y| \ge \epsilon |Y|
	\end{equation}
	where $\partial Y$ is the boundary of $Y$, which means the set of edges in $V$ that have one endpoint in $Y$ and one endpoint in $V \setminus Y$. 
\end{definition}

\begin{definition}
	(Isoperimetric constant) The isoperimetric constant of $G$, $h(G)$, indicates the minimum $\epsilon$ for the graph:
	\begin{equation}
		h(G) = \min \Big\{ \frac{|\partial Y|}{|Y|}  \Big\} 
	\end{equation}
	with $|Y| \le \tfrac{1}{2}|V|$ .
\end{definition}

The specific expander graphs we study here are the $k$-regular random graphs.

\begin{definition}
	($k$-regular graph) A graph $G$ is $k$-regular if every vertex has degree (valency) $k$. That is, every vertex connects to $k$ edges.
\end{definition}

\begin{definition}
	(Expander family, def 1.74 of \cite{expandersguide}) Let $k$ be a positive integer. Let $(G_n)$ be a sequence of $k$-regular graphs such that $|G_n| \rightarrow \infty$ as $n \rightarrow \infty$. $(G_n)$ is an expander family if the sequence $(h(G_n))$ is bounded away from zero.
\end{definition}

Let the eigenvalues of $G$ be $\lambda_0 \ge \lambda_1 \ge \dots \ge \lambda_{n-1}$. When $G$ is a $k$-regular graph, then $\lambda_0 = k$. This fact is easily demonstrated by noting that the eigenvector associated with $\lambda_0$ is $(1, 1, 1, \dots)\sqrt{n}$. The second-largest eigenvalue $\lambda_1$ is bounded by the Alon-Boppana theorem\cite{Alon1986, Nilli1991}. It can then be shown that the smaller $\lambda_1$ is, the larger $h(G)$ is,

\begin{proposition}
	(Prop 1.84 of \cite{expandersguide}) Let $G$ be a $k$-regular graph, then
	\begin{equation}
		\frac{k - \lambda_1}{2} \le h(G) \le \sqrt{2k(k - \lambda_1)}
	\end{equation}
\end{proposition}

Conversely, the larger $h(G)$ the larger the spectral gap $\lambda_0 - \lambda_1$ (recall that $\lambda_0 = k$). This is the reason that expander graphs are good candidates for emergent states. 

The isoperimetric constant is not so easy to calculate. It is closely related to the property of topological groups called weak containment, specifically a quantity called Kazhdan's property (T)\cite{KazhdanTbook}, which can sometimes be more easily bounded. However, for the present purposes, we simply point out that the expansion property of a family of graphs is best estimated from graph spectra. This is demonstrated for $k$-regular graphs by the Alon-Boppana bound\cite{Alon1986, Nilli1991}.

In recent work we have shown that $k$-regular random graphs retain their characteristic spectral properties even when a substantial fraction of edges have been removed at random\cite{ScholesEntropy}. We explain those numerical results on a more formal level, with proofs, in another paper. We found that the ensemble of graphs has the spectrum of an ensemble of $d$-regular random graphs, where $d \le k$ is the average valency of vertices for the ensemble of graphs. Therefore, these graphs can be quite disordered, but still display a prominent eigenvalue that is well-separated from those of the remaining `random' states. This particular point, and the reasons that the emergent state is spontaneously stable, will be developed in another paper.

In other work, it has been shown that complex networks where nodes are coupled with the structure of $k$-regular random graphs are highly synchronizable\cite{Townsend2023}. In that work the authors prove that for the case of regular graphs, spectral expansion implies global synchrony. That is, a network of phase oscillators coupled in the fashion of a $k$-regular graph will synchronize regardless of their initial phases. In the present work we do not necessarily need perfectly synchronizable networks. In other words, it may be advantageous to suffer some initial phase arrangements that inhibit synchronization with the benefit of less stringent construction requirements for the graph. After all, these networks will be much more robust to disorder and implementable if they can sustain a significant variance in structure. Our main requirement is that the graphs are on average $d$-regular.

Some examples of $k$-regular random graphs on $n = 40$ vertices, together with their spectra, are shown in Fig 2. Here $k = 20$, so the number of edges in the graph without edge disorder is $kn/2 = 400$. We plot edge-disordered graphs, with various numbers of edges removed. The deleted edges are indicated by green lines, whereas the retained edges are light blue. We compare cases where 20, 100 and 300 edges have been randomly removed. The emergent state---evident as the isolated, largest eigenvalue---is qualitatively stable (that is, separated from the rest of the spectrum) until more than half the edges have been deleted. The eigenvector of the emergent state is $(1, 1, 1, ...)/\sqrt{n}$.

\begin{figure}[!h]
	\centering\includegraphics[width=2.5in]{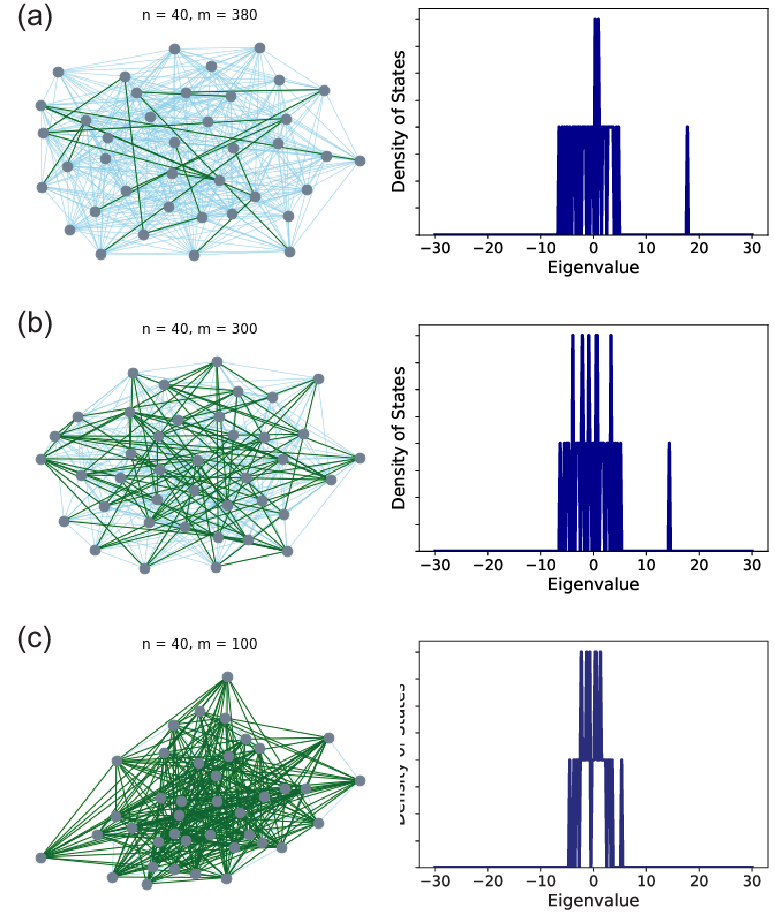}
	\caption{ Examples of $k$-regular random graphs on $n = 40$ vertices with $k = 20$, together with their spectra. See text. (a) The number of edges is $m = 380$ (note that the maximum possible number of edges is 400). Here the emergent state is clearly evident as the isolated state with greatest eigenvalue. (b) The number of edges is $m = 300$.  (c) The number of edges is $m = 100$. Here the average valency (degree) of the vertices is too small for the emergent state to separate clearly from the spectrum of random states. }
	\label{fig_2}
\end{figure}

It is useful to compare these graphs and spectra to those of signed graphs.  Let's say the largest eigenvalue of the unsigned graph is $d$, then, if the signs of all vertices are reversed, the smallest eigenvalue of the signed graph is $-d$. It is associated with the eigenvector $(1, 1, 1, ...)/\sqrt{n}$. We show in Fig 3 examples where some fraction of the edges are randomly given negative sign, according to probability $p_n$. The positive edges are blue, while the negative edges are red. We compare case of $p_n = 0.2$ to $p_n  =0.8$. It is evident that signed graphs also preserve the emergent state, but only when the number of edges with one sign is sufficiently greater than those of the other sign.

\begin{figure}[!h]
	\centering\includegraphics[width=2.5in]{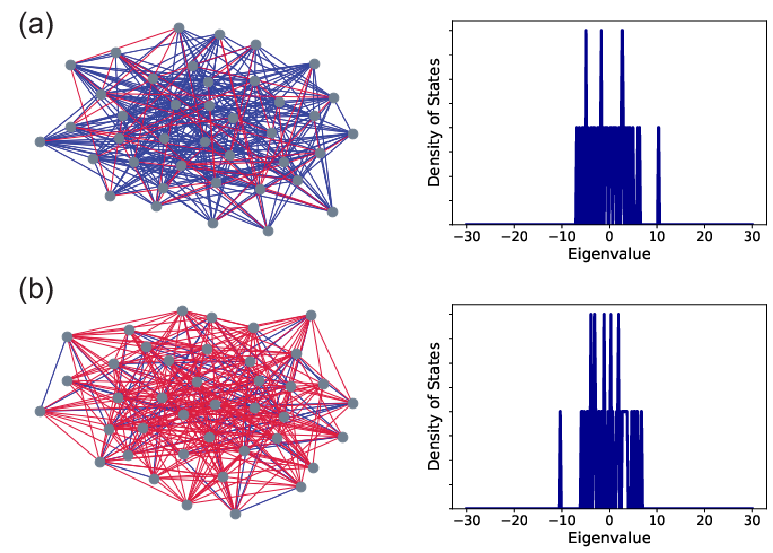}
	\caption{ Examples of $k$-regular random graphs with $n = 40$, $k = 20$, and $m = 300$ with signed edges. (a) The probability that en edge is negative is $p_n = 0.2$. Here, the positive edges dominate, so the emergent state has the greatest eigenvalue in the spectrum. (b) The probability that en edge is negative is $p_n  = 0.8$. Now, negative edges dominate, so the emergent state has the least (most negative) eigenvalue in the spectrum. }
	\label{fig_3}
\end{figure}

\section{Quantum-like states on networks}
Here we define properties that QL states should possess by giving some brief definitions of the state space we use for quantum systems and then proposing some axioms that QL states should satisfy. After that, we build concrete examples of systems that exhibit such states.

\subsection{Background and Axioms}
Our aim is to demonstrate stable states of a complex network that are defined in some basis, $\{ |a_1\rangle, |a_2\rangle   \}$, such that we can produce arbitrary states by arbitrary convex combinations of the basis states,
\begin{equation}
	|A\rangle = \alpha_1 |a_1\rangle +  \alpha_2 |a_2\rangle,
\end{equation}
where $\alpha_1, \alpha_2 \in \mathbb{C} \: \textrm{or} \: \mathbb{R}$ and by normalization $\alpha^2_1 + \alpha^2_2 = 1$. These vectors will comprise a Hilbert space $\mathcal{H}_A$.

It might be worth defining Hilbert space here, and for more detail see\cite{KadRing}. A Hilbert space $\mathcal{H}$ is a (complete) vector space over a field $\mathbb{K}$ (which can include $\mathbb{C} \: \textrm{or} \: \mathbb{R}$, for example) with an inner product $\langle x,y  \rangle$ which generalizes the idea of a dot product of vectors in Cartesian space. For any vectors $x, y, z$ in $\mathcal{H}$ and scalars $\alpha \in \mathbb{K}$ we have
\begin{eqnarray*}
	\langle x+y, z  \rangle = \langle x,z  \rangle + \langle y,z  \rangle \\
	\langle \alpha x,y  \rangle = \alpha \langle x,y  \rangle \\
	\langle x,y  \rangle = \overline{\langle y,x \rangle} \\
	\langle x,x  \rangle = \alpha \ge 0
\end{eqnarray*}
where the overbar means complex conjugate (or its equivalent for the relevant field). Note also that the `vectors' can be functions satisfying the properties listed above. The norm is defined in terms of the inner product
\begin{equation*}
	\lvert x \rvert =  \langle x,x  \rangle^{\frac{1}{2}} .
\end{equation*}

We will define in the following sections  QL states, constructed using networks  that have emergent states satisfying these properties. Here our vector spaces are defined explicitly over $\mathbb{R}$. If we want to define them over $\mathbb{C}$ we should use biased graphs\cite{Zaslavsky1, Reff2012, Mehatari}, with edges taking values from $\{ 0, 1, -1, i, -i \}$.

We propose that those QL states and/or their underlying graphs should satisfy the following axioms:

\begin{enumerate}
	\item The graphs will exhibit an emergent state that is distinguished in the eigenvalue spectrum.
	\item Each graph will be robust and stable to disorder in its precise construction as well as respect to disorder in the frequency of the oscillators represented by its nodes (vertices).
	\item The basis QL bit is a two-state system, but one of the two states is not `off' because the QL bit is defined by turning `on' a network of oscillators. Therefore we need the QL bit to comprise a basis of two distinguishable `on' states. What we mean by `on' and `off' distinguishes QL states from quantum states, where we are happy to have a two level system comprising the vacuum, or ground, state and another level. In the case of QL states the network needs to be activated (i.e. `on') to function as a network and to be able to produce superpositions by interacting with other networks of oscillating dipoles.
	\item Unitary operations on graphs and subgraphs should be defined.
	\item We need a construction that couples graphs or subgraphs in such a way that allows production of a state space homomorphic to superpositions of product states.
\end{enumerate}

\subsection{The quantum-like bit}
Khrennikov and co-workers initiated a program for identifying how classical systems could produce probabilistic outcomes like those peculiar to quantum mechanical systems. Progress has largely concerned how probability laws are modified, and how probability is measured, as we review in a later section. In the course of many examples they refer to suitable QL states. In one paper\cite{Khrennikov2018SciRep} a specific example is proposed for a single neuron. Here we take a different approach by exploring the possibility that QL states can be composed from interacting networks. A main motivation is that these states can be surprisingly robust to decoherence---whereas typical quantum states are likely to decohere quickly in disordered environments, which is the main objection to the idea that quantum phenomena play any functional role in biological systems. It is worth reiterating too, these states are not strictly quantum mechanical states; they are QL states. 

Building blocks proposed for QL states are networks, possibly very complex, that we define using graphs. Networks have been extensively studied\cite{Barabasi} and can have the properties of Axioms 1 and 2, as discussed earlier in this paper. Networks are known to provide insights into social networks, neural networks, biochemical and biological networks, communications systems, and so on. They can be functionally sophisticated and encode a lot of information. Networks can evolve through feedback to shift their balance and condense consensus. In short, these building blocks are rich and powerful structures, although in the present work we solely consider their emergent eigenstates.

The emergent states of  networks are classical. They arise from nonlinear interactions among the vertices so that synchronization---the key characteristic of the emergent state---arises from feedback\cite{Strogatz2000, Acebron2005}. This produces a stable eigenstate of the network of oscillators, which for $n$ oscillators is the state $(1, 1, 1, \dots)/\sqrt{n}$. Our task is to develop a way to build from these classical states, new states that have properties analogous key characteristics of quantum states. Specifically, we first need to define a QL bit $|A\rangle$ that `lives in' a Hilbert space $\mathcal{H}_A$.

It is proposed that QL states can be built from QL bits defined by partitioning an edge-disordered $k$-regular random graph $G$ into blocks, giving the subgraphs $G_{a_1}$ and $G_{a_2}$, and the tensor sum structure for the graph's adjacency matrix, $A = A_{a_1} \oplus  A_{a_2}$. We need to ensure that there is a lower probability of edges interconnecting the subgraphs than within each subgraph (for reasons discussed below). Thus we should delete some fraction of the interconnecting edges (those spanning $G_{a_1}$ and $G_{a_2}$). In the demonstrations below, we start from unconnected subgraphs (no edges between $G_{a_1}$ and $G_{a_2}$), then randomly add edges with some probability.

The edges within subgraphs, or those edges connecting subgraphs, can be positive or negative---that is, these are signed graphs. The physical interpretation of signed graphs is discussed below. The QL bit thus defined is a bipartite graph with respect to the vertex sets of the constituent pair of subgraphs $G_{a_1}$ and $G_{a_2}$. It has intrinsic $k$-regular random structure, ensuring that the emergent states are robust.

Some examples of QL bits are shown in Fig 4a, b. Here each subgraph is on 30 vertices, $k=20$, and the probability of an edge between any two vertices within a subgraph is approximately 0.23. The two subgraphs, $G_{a_1}$ and $G_{a_2}$ are shown coloured blue and green respectively. Coupling edges, that connect the subgraphs, are drawn in red. The left hand drawing shows the graph and all vertices, whereas the middle drawing shows the separated subgraphs, including all their edges, and the coupling edges are drawn schematically. In Fig 4a, the probability of adding a coupling edge between any two vertices, one in $G_{a_1}$ and the other in $G_{a_2}$, is 0.01. That probability is 0.1 in Fig 4b. In these examples, the edges within the subgraphs are all positive, whereas the coupling edges are negative. Note also that the value of the coupling edge (the coupling strength) is set to 1 because we are plotting the spectra of the adjacency matrices. This configuration makes the largest emergent eigenstate, shown in the spectra, $\frac{1}{\sqrt{2}}(a_1 - a_2)$ and the second largest $\frac{1}{\sqrt{2}}(a_1 + a_2)$, as indicated. Here, $a_i$ means the emergent eigenstate of subgraph $G_{a_i}$. The order of these eigenstates can be reversed by changing the signs of the coupling edges. 

The emergent superposition state can be tuned to a mixed state---that is, where it could be $a_1$ or $a_2$ with classical probability---by reducing the number of coupling edges (or their coupling strength) so that the subgraphs are not synchronized and the eigenvalues of $\frac{1}{\sqrt{2}}(a_1 - a_2)$ and  $\frac{1}{\sqrt{2}}(a_1 + a_2)$ converge. This is seen by comparing the graph spectra in Fig 4a and Fig 4b. Conversely, when the number of coupling edges is increased, the graph $G$ becomes overall $k$-regular; it is no longer described as a pair of coupled $k$-regular random subgraphs. Then what was in the other examples the second largest eigenvalue converges into the density of random states, Fig 4c. This is why we want to balance the density of edges within each of $G_{a_1}$ and $G_{a_2}$ relative to those coupling edges that connect the subgraphs. 

To summarize, the keys to the proposal of these QL bits are:
\begin{enumerate}
	\item By coupling a pair of coupled $k$-regular random subgraphs $G_{a_1}$ and $G_{a_2}$, we can produce convex combinations of the eigenstates of the subgraphs. These states therefore serve as basis states that can be generated as superposition states. \\
	\item The subgraphs are $k$-regular random graphs, which ensures that the basis states and their superpositions are very robust for networks of arbitrary complexity and are identified clearly by their emergent eigenvalue. \\
	\item The $k$-regular random subgraph construction also ensures that, regardless of the number of nodes (vertices) in each subgraph, the emergent eigenvalues are found at the same frequencies, that is, $d$, where $d$ is the average degree of vertices in the subgraph. $d$ is less than $k$ when the graphs are disordered and have randomly missing edges\cite{ScholesEntropy}. 
\end{enumerate}

\begin{figure}[!h]
	\centering\includegraphics[width=5.0in]{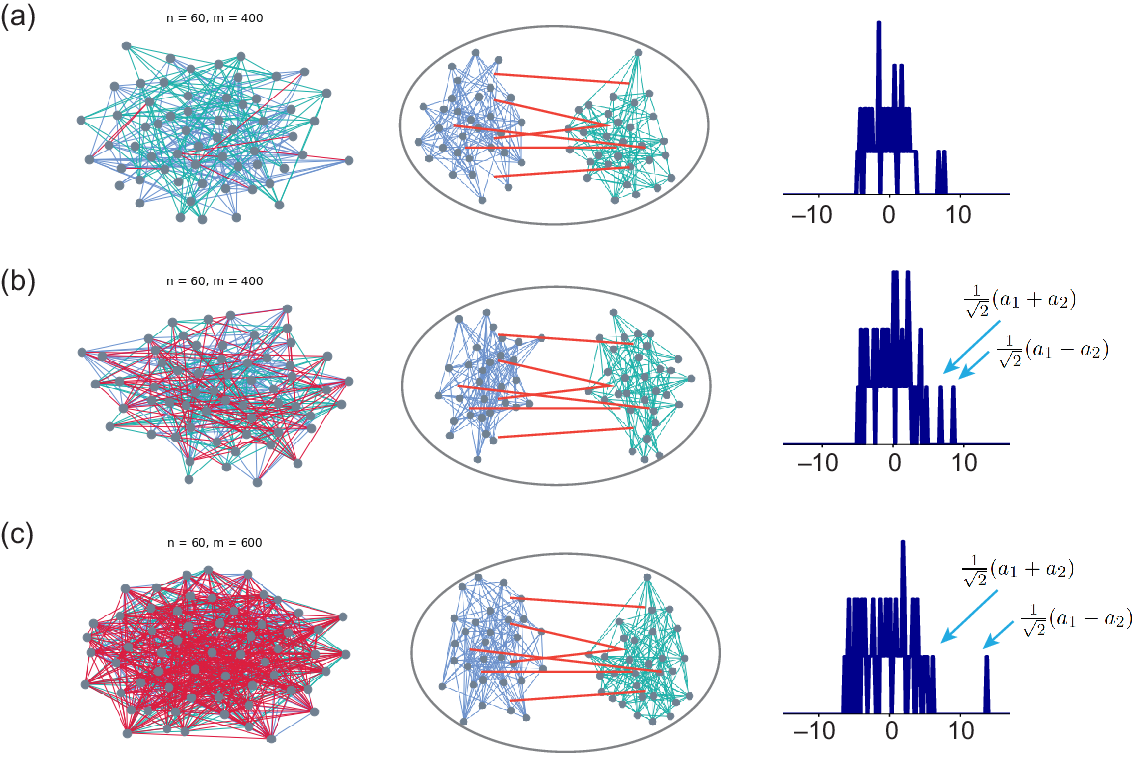}
	\caption{Examples of networks that produce QL states, as described in the text. The left diagrams show the network. The middle pictures shows the subgraphs $G_{a_1}$ and $G_{a_2}$ separated in space, and a schematic indications of the coupling edges between them. On the right the spectra of the graphs are shown. The eigenstates are written as linear combinations of the eigenstates of the subgraphs, that is, $a_1$ is the emergent eigenstate of $G_{a_1}$, and so on. In the graphs, the red edges are negative. Other edges are taken to be positive in these examples. }
	\label{fig_4}
\end{figure}

How do we think about what these graphs represent physically?  Our working model is that each vertex is an oscillator. The concept of an oscillator is very broad, but the the idea is that the vertex can flip periodically between two states or phases. It might be the phase of a neuron firing, it might represent the `for or against' an opinion given some information specific to a social sub-network, it might represent the phase of an electrical pulse in a circuit, or perhaps the concentration---above or below a threshold---of a biochemical marker. The coupling between vertices maps out interactions among these states. In this paper we describe the networks within each QL bit as perfectly correlated (all the edges within each subgraph have the same sign). That is not necessary for these constructions, as suggested by Fig 3. We simply need consensus within the network.

The QL bit can be synchronized to a superposition state of the constituent subgraphs, or it can display either subgraph probabilistically, thus satisfying Axiom 3. The physical concept of the graph is that it represents a network of oscillating dipoles, one positioned at each vertex. Edges indicate the dipole-dipole coupling between vertices. Negative edges denote coupling between oppositely-phased dipoles.

Unitary operations, including QL gates, can be imagined for these networks as graph homomorphisms. In particular, we simply need operations that flip subgraph or coupling edge phases. We will explain this, with examples, in a later section. The QL bits thus satisfy Axiom 4.

Lastly, these states are presented as spectra and associated states of graphs, that is, of the graphs' adjacency matrices. Therefore, the off-diagonal terms are all set to unity. In a realistic system, these off-diagonal terms have energy values reflecting the coupling and the diagonal entries can have frequency disorder\cite{Scholes2020}. Here we do not need to examine various parameters because we are attempting to establish `existence' of the prescribed QL states---are they feasible with the right kind of network? The key to the argument in favour is that the eigenvalue gap between the emergent state and the next state is fixed by the mean valency of the disordered $k$-regular graph, $d$. For large networks, then, we simply need to ensure that $d$ is `large enough' so that the QL-bit superposition state is emergent, that is, distinct in the spectrum.

\subsection{Quantum-like states}
Two or more QL bits can be suitably coupled so that QL states are constructed. By imagining the QL bit as an oscillating dipole (or transition density\cite{Mirkovic2017, Scholes2003}), then we can propose how `wiring diagrams' can lock the relative phases of subgraphs of each QL bit in lock-step across the multi-QL bit network. Then the coupling between two QL bits labelled A and B, with basis states $a_1, a_2$ and $b_1, b_2$ respectively, can be thought of, conceptually, as a dipole-dipole coupling\cite{Scholes2003}. The coupling Hamiltonian is given by:
\begin{equation}
	H = \hbar \nu_A |A\rangle \langle A | + \hbar \nu_B |B\rangle \langle B | + \sigma J(|A\rangle \langle B |  + |B\rangle \langle A | )
\end{equation}
where $\nu_A$ is the oscillation frequency of QL bit A, $\nu_B$ is the oscillation frequency of QL bit B,  and $J$ is the coupling.  We have $\sigma = \pm 1$ depending on whether the QL bit pairs are coupled so that the oscillations $a_1 \rightarrow a_2$ and $b_1 \rightarrow b_2$ are in-phase or anti-phased, respectively. In our calculations of the the graph adjacency matrices, we have $\nu_A = \nu_B = 0$ and $J = 1$. The basis states are emergent states of the subgraphs, for example by $a_1$ we mean the emergent state of subgraph $G_{a_1}$.

If we think of each QL bit conceptually as an oscillator in a superposition of states of its subgraphs $|x_1\rangle$ and $|x_2\rangle$, where $x \in \{ a, b, c, \dots\}$, then we can envision how suitable connections among  QL bits can produce a variety of synchronized states reminiscent of the set of states in $\mathcal{H}_A \otimes \mathcal{H}_B \otimes \mathcal{H}_C \otimes \dots$. To do this, we introduce signed edge sets that lock subgraphs of the QL bits in or out of phase with each other, so that the entire graph is locked relative to a global phase.  

What we need to establish is a representation of the basis of the relevant tensor product of Hilbert spaces as a pattern of phases (relative phases) of the eigenstates of network subgraphs in the emergent eigenstate of the entire network of coupled QL-bits.  The product basis for quantum states has inbuilt correlations that we can represent as linear combinations in a free vector space that will serve as the QL representation, physically laying out the phase correlations across the network. This is a key technical challenge that we will address in future work and highlight some open questions. While the method proposed here enables us to produce superpositions of states within any excitation subspace, generating more complex states requires further work. The idea we use is that a set of appropriate linear combinations of QL-bit states (naturally produced by joining QL-bits by edges) allows us to construct a QL state. 


Here we only consider the Bell states in order to explain the concept, Fig 5. Fig 5a shows how the superposition states are produced by coupling the QL bits. Any of these four states can be selected as the emergent state by suitable choice of signs of the couplings within each QL bit and between the QL bits. The case of positive-signed edges between the QL bits is shown here. The phases of the blocks of coefficients in the resulting eigenstates are indicated, with ordering $a_1$, $a_2$, $b_1$, $b_2$. In Fig 5b, c we show schematics of how to construct networks that emulate two of the Bell states, $| \Psi_-\rangle$ and $| \Psi_+\rangle$. The pictures of the networks shows the idea of how the coupling between subgraphs controls their relative phases. The analogy of coupled dipoles shows the same concept and the phase map sketches indicate how the subgraph phases are phase-locked to each other. Calculated spectra are shown in the other panels, Fig 5d-g. First, for reference, in Fig 5d we show the spectrum of $G_{a_1} \oplus G_{a_2} \oplus G_{b_1} \oplus G_{b_2}$, where none of the subgraphs are coupled. This is the perfectly mixed state. In Fig 5e, the spectrum of the network where the subgraphs within each QL-bit are coupled (with random edges between subgraphs within each QL-bit added with probability $p = 0.2$), but the QL-bits are not coupled to each other. So, here we have QL-bits but no entangled states. The adjacency matrix is therefore represented as two blocks, $G_A \oplus G_B$.

In Fig f, g we show spectra where QL-bits are coupled, removing the block form of the adjacency matrix, to produce an emergent state corresponding to the Bell state $|\Psi_- \rangle$. The inter-QL-bit coupling edges are added with $p = 0.05$ and with signs as indicated in Fig 5b. In Fig 5g we show the spectrum when all coupling edges are added with $p = 0.05$, that is, both the intra-QL-bit and inter-QL-bit edges. For each calculation, the composition of the emergent eigenstate is indicated. 

\begin{figure}[!h]
	\centering\includegraphics[width=5.0in]{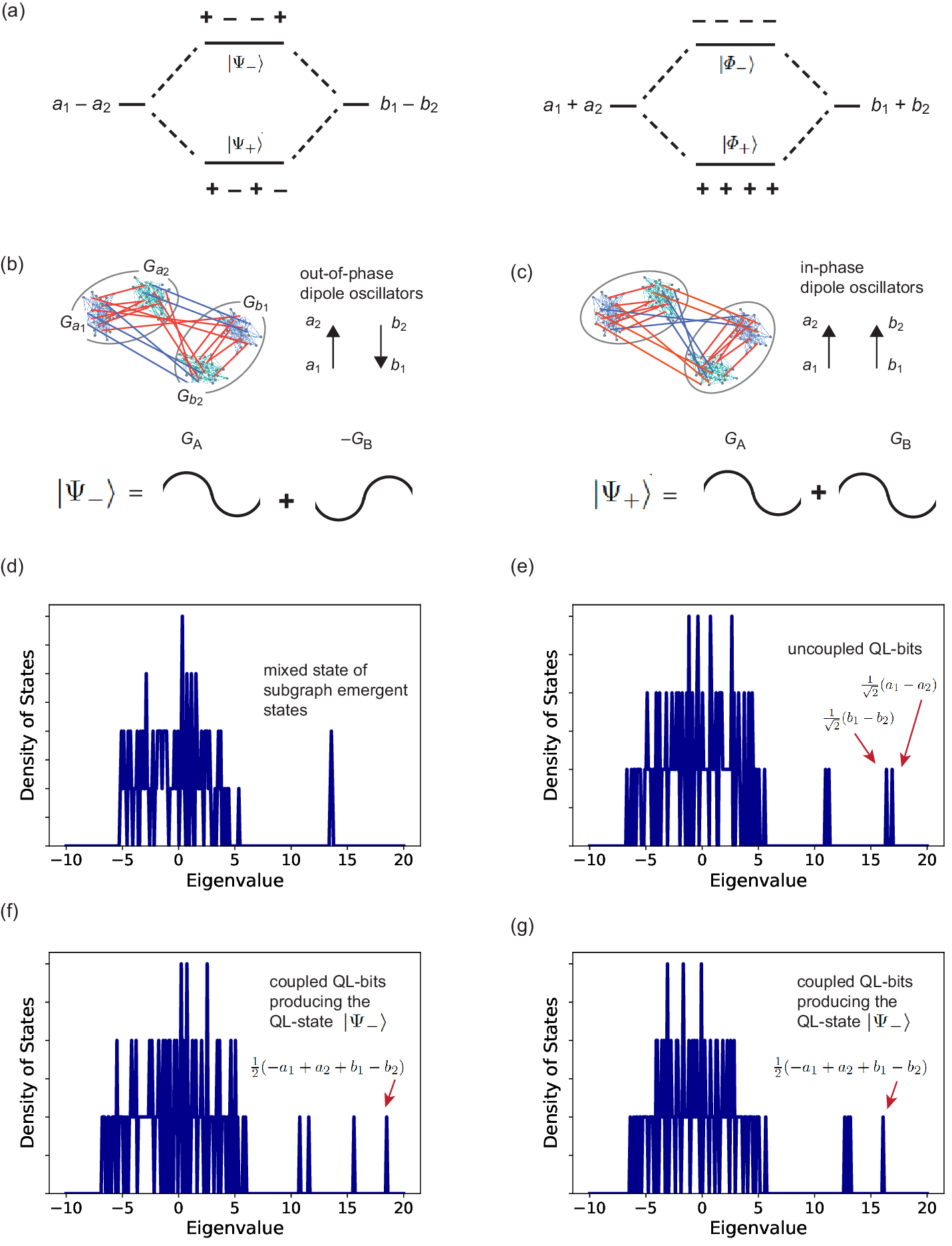}
	\caption{(a) The superposition states are produced by coupling the QL bits. Here constructions of the four Bell states are shown. (b, c) Diagrams that show how the QL-bit is analogous to a pair of coupled oscillating dipoles. To change the state of the QL-bit from out-of-phase oscillations to in-phase oscillations, the signs of the edges need to be reversed, as indicated. Red edges are negative, blue edges are positive. The corresponding phase map sketches are also shown for reference. (d) The spectrum of $G_{a_1} \oplus G_{a_2} \oplus G_{b_1} \oplus G_{b_2}$, that is none of the subgraphs are coupled. (e) The spectrum of the direct sum of QL-bit graphs, that is, the subgraphs within each QL-bit are coupled (with random edge added with probability $p = 0.2$), but the QL-bits are not coupled to each other. (f) The QL-bits are coupled to produce an emergent state corresponding to the Bell state $|\Psi_- \rangle$. The coupling edges are added with $p = 0.05$ and with signs as indicated in part (b). (g) The QL-bits are coupled to produce an emergent state corresponding to the Bell state $|\Psi_- \rangle$, like in part (e). The difference in this calculation \emph{all} coupling edges are added with $p = 0.05$, see text. }
	\label{fig_5}
\end{figure}

We can characterize the resulting states according to the overall `excitation number' of the $|x_2\rangle$ states, that is, just like the excitation subspaces of $\mathcal{H}_A \otimes \mathcal{H}_B \otimes \mathcal{H}_C \otimes \dots$. Synchronization locks the QL bit array into a particular global state (which can also be a mixed state, caused by insufficient synchronization).  Just like in the coupled dipole analogy, when $N$ QL bits are coupled, we can specify $N$ states that map to superpositions of product states comprising one $|x_2\rangle$ state and $N-1$ of the $|x_1\rangle$ states---the single excitation subspace. We have ${N \choose 2}$  double excitation states, and so on. A frequency difference between $|x_2\rangle$ and $|x_1\rangle$ is allowed, but not essential, depending on how the states are detected.

The QL bits are networks that can be spatially separated from each other (like shown in Fig 5), or they can be completely entwined. They can be different sizes (different numbers of nodes) and can be slightly detuned from one another (thus changing how the states mix). There is a rich variety of ways that the system can be composed. The important rule is that the expander subgraphs are approximately $d$-regular, so that the eigenvalues of each QL bit are close to resonance with the others. 

\section{Context and the V\"{a}xj\"{o}  model}
A model for QL measurements has been developed to apply to a wide range of classical systems, the V\"{a}xj\"{o} model\cite{Khrennikov1}. These developments complement our definition of QL states, which give concrete examples of how to engineer or identify complex systems that yield quantum-like probability laws.

The V\"{a}xj\"{o} model does not require specification of systems and their states. Instead, it focuses on measurement and probability laws. Specifically, how probability laws might be formulated so that interference-like phenomena are accounted for. By appending our observables with extra structure that relates to conditions of the measurement and other information potentially gathered through prior measurements---context---Khrennikov shows how probabilistic laws can incorporate interference effects. He defines the probability for obtaining $a = \alpha$ by observing $a$ under context $\mathcal{C}$ as:
\begin{equation}
	p^a_{\mathcal{C}}(\alpha) \equiv \mathbb{P}(a = \alpha | \mathcal{C}),
\end{equation}
where each contextual probability satisfies $p^a_{\mathcal{C}}(\alpha) \geq 0$ and
\begin{equation}
	p^a_{\mathcal{C}}(\alpha_1) + \dots + p^a_{\mathcal{C}}(\alpha_n) + \dots = 1
\end{equation}
for any context $\mathcal{C}$.

\begin{definition}
	(Khrennikov) Contextual expectation $\mathbb{E}[a|\mathcal{C}]$ of an observable $a \in \mathcal{O}$ with respect to context $\mathcal{C}$ is given by
	\begin{equation}
		\bar{a}_{\mathcal{C}} = \mathbb{E}[a|\mathcal{C}] = \alpha_1 p^a_{\mathcal{C}}(\alpha_1) + \dots + \alpha_n p^a_{\mathcal{C}}(\alpha_n) + \dots  .
	\end{equation}
\end{definition}

\begin{definition}
	(V\"{a}xj\"{o} model) A contextual probability model is a contextual probability space $\mathcal{P} = (\mathcal{C}, \mathcal{O}, \pi)$ such that $\mathcal{C}$ contains a special subfamily of contexts $\{ \mathcal{C}^a_{\alpha} \}_{a \in \mathcal{O}, \alpha \in X_a}$ which are interpreted as $[a = \alpha] $-selection contexts. Context $\mathcal{C}^a_{\alpha}$ corresponds to the selection with respect to the result $a = \alpha$. Moreover, it is assumed that $\mathcal{C}^a_{\alpha}$ satisfies the condition
	\begin{equation}
		\mathbb{P}(a = \alpha | \mathcal{C}^a_{\alpha} ) = 1.
	\end{equation}
	It is assumed that, for each observable $a \in \mathcal{O}$ and its value $\alpha$, the selection context $\mathcal{C}^a_{\alpha}$ is uniquely determined in the class of contexts $\mathcal{C}$ of the model.
\end{definition}

We also need to define what Khrennikov terms `transition probabilities' using a conditional probability construction:
\begin{equation}
	p_{\beta | \alpha} \equiv \mathbb{P}(b = \beta | a = \alpha) = \mathbb{P}(b = \beta | \mathcal{C}^a_{\alpha}),
\end{equation}
because QL probability laws will be evidenced by measurements on two observables, $a$ and $b$. The reader is referred to \cite{Khrennikov1} for more details. 

The formula for total probability in the usual, Kolmogorov, model (i.e. without the structure of contexts) is
\begin{equation}
	p^b(\beta) = \sum_{\alpha} p^a(\alpha) p_{\beta | \alpha}.
\end{equation}
But, as Khrennikov argues, this formula is not guaranteed to hold in the V\"{a}xj\"{o} model, where probability distributions in the set of data for $a, b, \mathcal{C}$ may not be described by a single probability space. For instance, the context for observation $a$ might vary depending on whether we previously learned something about $b$, through observation (or \textit{vice versa}). That concept is exploited by the `question order effect', discussed below.

The deviation of the contextual total probability from the Kolmogorov law serves as a kind of interference term, $\delta(\beta | a, \mathcal{C})$:
\begin{equation}
	\mathbb{P}(b = \beta | \mathcal{C}) \equiv p^b_{\mathcal{C}}(\beta) = \sum_{\alpha} p^a_{\mathcal{C}}(\alpha)p_{\beta | \alpha} + \delta(\beta | a, \mathcal{C}).
\end{equation}
The interference term is the key to connection with quantum systems, where interference of probability amplitudes is a defining feature. The interference term identified here is the classical counterpart to conditional expectation in quantum theory\cite{Ozawa1985}.

The QL states of classical synchronized systems proposed in the previous section can represent states encoded with context. That context essentially means that the QL bit networks or collective QL states respond and change as information is gathered by other QL bits---that is, we make measurements. This could be accomplished in various ways by connecting QL bits and/or adding additional `context' networks.

\section{Examples of quantum-like function}
Here we provide two simple, but very different, examples of how QL bits might serve as a the basis for information processing. The first example integrates the QL bits into Khrennikov's concept of the question order effect, that exploits interference effects that come from context, reflecting the order that information is gathered. It indicates how to think about measurement of QL states. The second example is a textbook protocol that exploits entanglement. It provides a reference point for assessing QL state functions.

It is important to note that quantum states and QL-states do not necessarily enable functions that are impossible to achieve by classical processes, but the different mechanisms that they allow to achieve the function could have advantages. The examples below illustrate different mechanisms of quantum functions.

\subsection{The Question Order Effect}
Khrennikov and co-workers have studied how QL measurement theory can be useful for explaining psychological measurements\cite{KB2013, Ozawa2020, Khrennikov2016}. That work provides an example of how the action of measurement matters when we are dealing with QL states. A prototypical example is the `question order effect'. The question order effect refers to the statistical observation that the answers to questions, for example in a survey, can depend on the order in which the questions are presented. Here we describe the QL measurement model for the question order effect qualitatively, using Figure 6. We refer the reader to the paper by Ozawa and Khrennikov for the details\cite{Ozawa2020}. 

\begin{figure}[!h]
	\centering\includegraphics[width=5.0in]{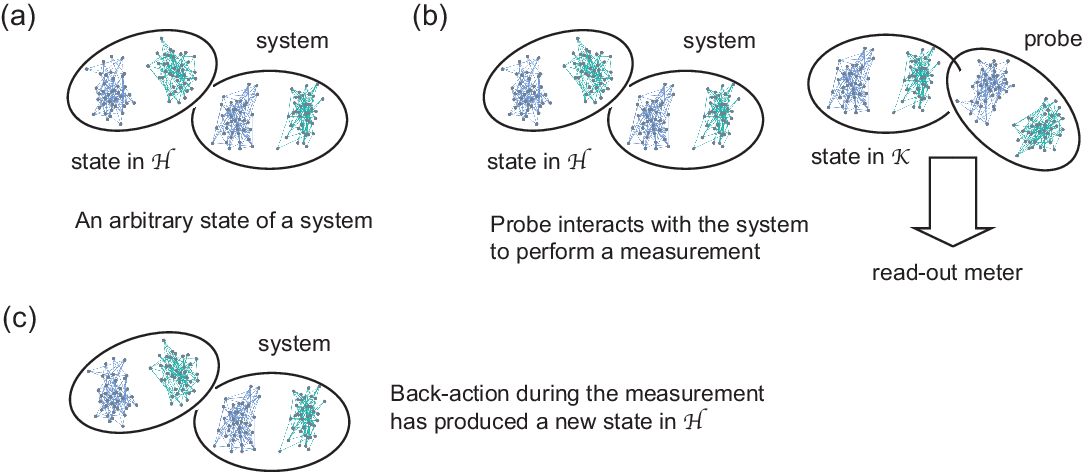}
	\caption{Diagram that shows the way networks interact for the steps needed to understand the question order effect described in the text.  }
	\label{fig_6}
\end{figure}

The system comprises a pair of `entangled' QL states. These projections $A$ and $B$ on the Hilbert space of the system, $\mathcal{H}$, correspond to questions labelled $A$ and $B$ respectively. For each question we can think about eigenvalues corresponding to answers of yes or no.  To measure the quantum state of our system, for example to measure the answer to question A, we need to interact the system with a measurement probe. This probe also comprises an entangled pair of QL states with projections on the Hilbert space $\mathcal{K}$. During measurement of $A$, for example, the composite system evolves according to a unitary operator $U_A$ acting on $\mathcal{H} \otimes \mathcal{K}$. The interaction encodes the probe with the eigenvalue of $A$ that can be read out using a meter observable. The interaction concomitantly changes the state of the system, so that a subsequent measurement on $B$ can give a different answer than if this measurement were carried out before the measurement on $B$. This interplay between system and measurement can model the question order effect.

The example shows how an appropriate `read out' of states is important for the kinds of sophisticated functions enabled by networks, in particular when a sequence of evolving answers or iterations may be needed. A good example is interactions among the states to reach `consensus'.

\subsection{Superdense Coding for Sensing or Processing}
Superdense coding is a strategy whereby  a quantum resource (entanglement) can mediate communication of two classical bits of information, potentially over a long distance, In an illustrative example, we  send two classical bits (00, 10, 01, or 11) using a single qubit\cite{NielsenChuang}.

Two QL bits can be `entangled', that is, synchronized, to produce a reference Bell-like state. Now apply a quantum gate to QL-bit A in isolation. In practice, this means we switch phases appropriately of the subgraphs in the QL network A. This can be understood intuitively using the analogy where each QL bit comprises a pair of coupled oscillating dipoles. In practice, we simply need to flip the signs of the edges within subgraphs and/or of the coupling edges. The details for each quantum gate are shown in Fig 7. Here we use the usual notation\cite{ScholesQIS} for the states and name QL-bit A Alice and QL-bit B Bob. We can associate $|0\rangle_A$ with $|a_1\rangle$ and $|1\rangle_A$ with $|a_2\rangle$, etc.

Bell-like states correspond to the four unique ways the two pairs of oscillating dipoles are locked in sync, Fig 7. Relative to the fixed phases of QL bit B, then, we can rotate the phases in QL bit A.  The information sent to QL bit B amounts to a single QL bit. However, a readout of the total Bell-like state of the A-B QL-bit pair translates into two classical bits of information. Two qubits are needed to encode two classical bits. Therefore we have the same total information supplied by four Bell states as four classical pair states. However, by making use of entanglement, QL bit A can communicate a pair of classical bits by sending only a single qubit. Notice that Alice's QL-bit has four possible states, but pairs of them differ only by an overall phase, so they do not convey different information in isolation. It is only with the context of the QL state that the four QL-bit states translate to four different interpretations. The key is that context is provided by referencing phase mutually among the QL-bits.

\begin{figure}[!h]
	\centering\includegraphics[width=5.0in]{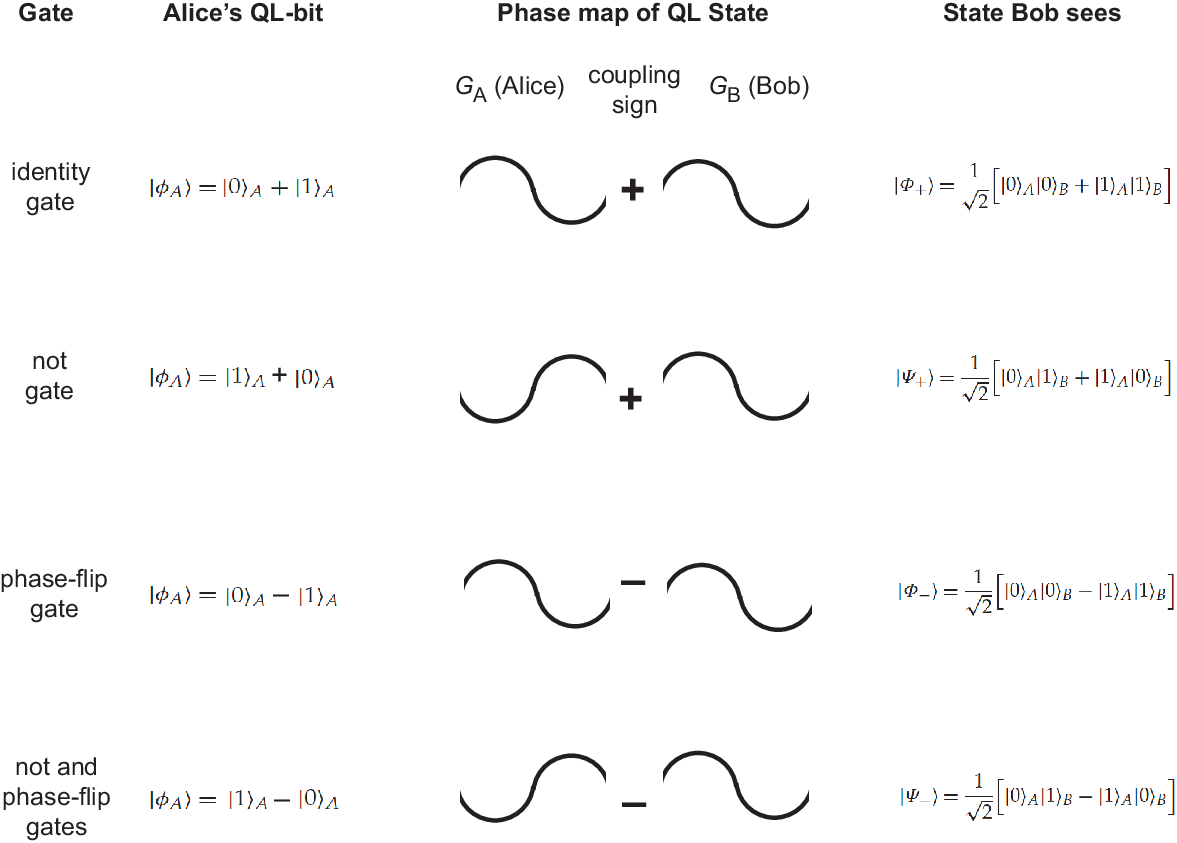}
	\caption{ A table that shows how Alice rotates the state of her QL-bit by a series of quantum gates and how that changes the QL state observed by Bob. The operations within the networks that produce the QL states are indicate schematically as a` phase map'. }
	\label{fig_7}
\end{figure}

The total Bell-like state of the A-B QL-bit pair could be detected by its spectrum. It is not obvious that the scheme would be useful for network communication when the entangled network states are close together, because we cannot exploit `spooky action at a distance'. However, the protocol might allow QL-bit A to serve as a quantum sensor, with read-out facilitated by QL-bit B. The proof of principle may also inspire ways that QL circuits could be constructed for more sophisticated processing operations.

A main conclusion from this example is that it is feasible to implement various quantum-like gates on the network states. For instance, Hadamard gate that `rotates' $| a_1 \rangle$  to the superposition $(| a_1 \rangle + | a_2 \rangle)/\sqrt{2}$ could be achieved by activating the coupling edges between subgraphs and activating the network corresponding to $G_{a_2}$. Then the state with the frequency of the emergent state---recall that is fixed by the valency of the $k$-regular graphs---is becomes the required superposition. These gates are quantum \emph{like} because the states and gates are composed by dynamic structures of the network, not by matrix operations. Nevertheless, then set of ways the network can be reconfigured offers a representation for the set of matrix operations that, in turn, represent maps on states in the relevant Hilbert space.

\section{Discussion}
In this study we have asked: Is it worth looking for quantum effects in very complex, ‘messy’ systems? The standard answer is that quantum states are too fragile to exist in such systems, including biological settings. However, given the many relevant new ideas that have emerged over the past decades—including, for instance, more sophisticated understandings of networks, synchronization phenomena, and advances in solving quantum dynamics for large and complicated systems, we argue that the time is right to revisit this question. Furthermore, perhaps we do not need true `quantum states', but instead sufficiently `quantum like' (QL) states will suffice. That inspires a shift from the common idea that we need to make chemical systems pristine in order to host quantum states in the form of qubits, to the idea that there are new opportunities to discover by identifying more abstract, but surprisingly robust  states. These states can even be realized in electronic  circuits. The main objective of this work is to propose concrete examples of very large and complex networks that host states resilient to disorder and displaying QL properties. 

By considering suitably constructed graphs with expander properties, it turns out that we can find robust QL states that can inhabit huge, complex systems and that satisfy the axioms proposed in Sec. 3. A requirement for states of any large and complex quantum system, whose structure is, in a sense, ‘organic’, is that they must especially stable with respect to various kinds of disorder. This is where the concept of emergent or collective phenomena is  important because it allows the state of interest to be distinguished from the many other states in the spectrum. This emergence property thus protects the state. To abstract a suitable model we need to account for an underlying structure, that could be encoded by some kind of rule (here the rule is that the graph is $k$-regular). Then we allow disorder in that structure or on the structure so that there is a large margin for flexibility in the precise structure. Frequency disorder at sites in a network is an example of disorder \emph{on} the structure\cite{Scholes2020}. Disorder caused by randomly deleting edges from the network is an example of disorder \emph{in} the structure\cite{ScholesEntropy}. We explicitly included this latter kind of disorder in the examples exhibited in the present work.

We identified $k$-regular graphs as examples of networks that can serve as the basis for QL states. However, any expander graph would serve this purpose. The best possible expanders are Ramanujan graphs\cite{Sarnak1988, Ramanujangraphs, Lubotzky2019}, but ideal expansion properties do not appear to be crucial because in many interesting or useful `real world' scenarios, it is desirable for the graphs to be disordered---so that precision construction is not needed. The expander graphs also have intrinsically interesting properties that are desirable for networks involved in communication and information transfer and processing\cite{Expanders, networks1, networks2}. Expanders are not necessary for the constructions, just sufficient. Other graphs might be of interest, including chiral graphs that could host states analogous to electron spins or photon polarization.

By positing a QL bit structure based on coupled expander graphs and combining that with the concept of QL states, here an ‘existence proposal’ for huge quantum states was described. The work opens up many questions for future work. Below we discuss two open questions: (i) Is there a `quantum-like' advantage for function? (ii) Could QL states and function be used by neural networks in the brain?

\subsection{Is there a QL advantage?}
Synchronized networks of QL states could serve as a hardware platform for information processing. There are three big questions: (i) what is the QL advantage for certain functions? (ii) What platforms could be designed to produce and test QL states? (iii) What experiments could test the QL functions of QL states?

The main point argued in the present work is that we defined construction of a QL bit $|A\rangle$ so that its emergent states `live in' a Hilbert space $\mathcal{H}_A$, just like quantum states.  A set of axioms is proposed that guide the properties the states need to have. We exhibited examples of these states and the way they are measured. It was also shown how QL gates can be implemented by controlling phases---specifically, signs of edges within each QL bit. Based on this framework, one can in principle adapt circuit designs from quantum processing systems. This is a key direction for future work that will provide answers to how QL states can have a QL advantage for function.

Compelling demonstration of QL state function requires a hardware platform for producing those states. The obvious approach is to design circuits of coupled electronic oscillators\cite{osc-circuit}. These kinds of systems could be miniaturized, and may be well suited for tasks like pattern recognition\cite{nanotech}. QL states might also be implemented in the framework of spiking neural networks, thereby taking advantage of neuromorphic hardware systems\cite{spiking1}.

Later work can adapt the concepts for more complex networks in optical, biological, chemical, and other systems. Biological systems are certainly an intriguing target because they comprise a huge number and diversity of networks. Manytimes these networks are so dense and complicated that they are referred to as `hairballs'\cite{hairballs}. The QL-states proposed here provide a platform for re-thinking quantum biology because it is much more likely to find QL-states that quantum states in such complex systems.

\subsection{Quantum neural networks}
It is well-known that neurons are connected to form the complex networks of the brain. It is now generally accepted that the ways these networks process signals and information to produce responses and functions do not simply involve signal transmission. Rather, information can be encoded in the timing and spatial coherence of signal oscillations\cite{memory, brain1, brain2, brain3, brain4, brain5, brain6, brain7}. We are not attempting to review that field here, but simply to sketch a small part of what is known to give examples of how networks function in what is arguably nature's most sophisticated machine---the brain. 

One key discovery is that information processing seems to be facilitated by oscillations of neuron activity. There are characteristic frequencies for these oscillations, grouped into bands denoted delta (<4 Hz), theta (4–8 Hz), alpha (8–12 Hz), and beta (12-30 Hz), in the low frequency range, to the high frequencies in the gamma band (30–80 Hz) and above. 

One  example is shown in Fig 8. Here gamma band oscillations are studied in the visual cortex of a monkey as it was presented with an image\cite{monkey1}. In the experiment, neuronal activity was detected using an electrocorticographic grid that monitored the left brain hemisphere of a macaque monkey. A number of prior studies had shown that neuronal gamma-band synchronization in visual cortex is associated with various kinds of functions. This study suggested that 50 to 80 Hz gamma-band activity plays a key role in the process of viewing. Fig 8 shows the main result of the paper, a unipolar local field potential trace taken as the monkey looks at two oranges. 

\begin{figure}[!h]
	\centering\includegraphics[width=2.5in]{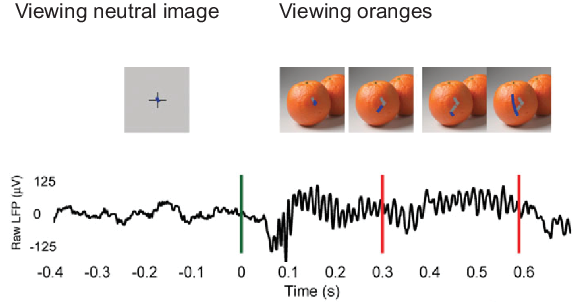}
	\caption{Gamma band oscillations observed in the visual cortex of a monkey as it was presented with an image. The monkey's focus is tracked by the line on the image. Reproduced with permission from \cite{monkey1}. }
	\label{fig_8}
\end{figure}

A second example serves to illustrate that brain function involves the interaction of widely distributed cortical regions. Long-range synchronization has been implicated in enabling the coordination of these regions across the brain and to communicate between brain regions\cite{brain4}. For example, how do we map sound to meaning\cite{brain3}? In \cite{corticalnetworks} the authors study the perception process. They used an analysis of EEG recordings in humans that allowed correlation of brain oscillations to be mapped. The experiment involved the subject watching a screen where two bars come together, then move apart. There is an audio cue when the bars meet. The measured power response during the stimulus is collected in maps, depicting the theta (4–8 Hz), alpha (8–16 Hz), beta (16–32 Hz), low gamma (32–64 Hz), and high gamma (64–128 Hz) band activity across the brain. The main finding of the work was enhanced long-range beta-synchrony during stimulus processing, which was accompanied by local suppression beta-band activity.

There is an immense body of work exploring brain function. A key question is how different sensory inputs are linked to appropriate processing circuits to draw conclusions. It has been proposed that `binding' of these units is facilitated by phase synchronization\cite{brain1}. It has also been proposed that the oscillation frequencies allow these disparate computations to associate\cite{brain5}. Further ideas hypothesize that a complex interplay of oscillations is important for perception\cite{brain7}. 

In sum, on one hand the brain functions using a complex network of networks that become synchronized during function. Here is where there is compelling evidence that coherent oscillations on complex networks are functionally operative in the brain, so a  connection with the ideas hypothesized in present paper---that QL function is possible---is feasible. However, it remains an enticing idea that requires further work. In particular, the way the dynamic networks interplay in the brain, and even how coherence is exploited are questions that have not yet been elucidated in detail. The present work at least provides a framework for devising testable hypotheses for additional, likely surprising, mechanisms that might be at play in the sophisticated networks of the brain. 

\section{Conclusion}
We proposed a quantum-like (QL) bit structure based on coupled expander graphs. Combining that with the concept of QL states, we suggest a way that huge, complex systems can host states that can process information in a QL fashion.

Specifically, it was suggested that building blocks suited for QL states are networks, possibly very complex, that we defined using graphs. These networks can be functionally sophisticated and encode a lot of information that is distilled into the emergent states we can exploit for QL like processing. The emergent states of  networks are classical. They arise from feedback. We developed a way to build from these classical states, new states that have properties analogous to quantum states. Specifically, we based this construction on a QL bit that `lives in' a Hilbert space $\mathcal{H}_A$. We demonstrate a candidate QL bit by partitioning an edge-disordered $k$-regular random graph $G_A$ into two blocks, giving the subgraphs $G_{a_1}$ and $G_{a_2}$, then connect those blocks to yield superposition states of the emergent states of each subgraph. The intrinsic $k$-regular random structure ensures that the emergent states are robust.

The work opens up many questions for future work. We explicitly discussed two major open questions: (i) Is there a `quantum-like' advantage for function? Perhaps this could be leveraged in hardware for QL-computing, which would be greatly simplified from that needed for true quantum computing, but may offer an attractive compromise between classical and quantum architectures. (ii) Could QL states and function be used by neural networks in the brain? We conclude that quantum-like states can exist in arbitrarily complex systems and that it is worthwhile investigating further examples and working out where they can have a functional advantage. The primary advance of the paper was to overcome the main objection for various proposals of quantum states in biology, the brain, and other complex systems---that is, that quantum states are too fragile to serve a functional role. This was done by proposing, first that we should think about `quantum like' states\cite{Khrennikov1}, not quantum states. Second, that we can devise a blueprint for suitable states, satisfying the axioms proposed above, by exploiting expander properties of graphs so that complex networks yield a single key state in their spectrum.   \vskip6pt

\section*{Acknowledgments}
This material is based upon work supported by the National Science Foundation under Grant No. 2211326 and the Gordon and Betty Moore Foundation through Grant GBMF7114. Graziano Amati and Andrei Khrennikov are thanked for helpful suggestions for improving the paper.

\bibliographystyle{unsrt}  
\bibliography{Scholes_4}

\begin{thebibliography}{10}

\bibitem{NielsenChuang}
M.~A. Nielsen and I.~L. Chuang.
\newblock {\em Quantum Computation and Quantum Information}.
\newblock Cambridge University Press, Cambridge, 2016.

\bibitem{Zurek2003}
W.~H. Zurek.
\newblock Decoherence, einselection, and the quantum origins of the classical.
\newblock {\em Rev. Mod. Phys.}, 75:715--775, 2003.

\bibitem{shadows}
R.~Penrose.
\newblock {\em Shadows of the Mind: A Search for the Missing Science of
  Consciousness}.
\newblock Oxford University Press, Oxford, 1994.

\bibitem{ScholesQIS}
Gregory~D. Scholes.
\newblock A molecular perspective on quantum information.
\newblock {\em Proc. R. Soc. A}, 479:20230599, 2023.

\bibitem{Khrennikov1}
A.~Khrennikov.
\newblock {\em Ubiquitous Quantum Structure}.
\newblock Springer, Heidelberg, 2010.

\bibitem{ScholesEntropy}
G.~D. Scholes.
\newblock Large coherent states formed from disordered k-regularrandom graphs.
\newblock {\em Entropy}, 25:1519, 2023.

\bibitem{Fano1992}
U.~Fano.
\newblock A common mechanism of collective phenomena.
\newblock {\em Rev. Mod. Phys.}, 64:313--319, 1992.

\bibitem{StrogatzBook}
S.~Strogatz.
\newblock {\em Sync: How order emerges from chaos in the Universe, Nature, and
  Daily Life}.
\newblock Hyperion, New York, 2003.

\bibitem{Strogatz2000}
S.~H. Strogatz.
\newblock From kuramoto to crawford: exploring the onset of synchronization in
  populations of coupled oscillators.
\newblock {\em Physica D}, 143:1--20, 2000.

\bibitem{Acebron2005}
J.~Acebrón, L~Bonilla, C~Pérez-Vicente, F~Ritort, and R~Spigler.
\newblock The kuramoto model: a simple paradigm for synchronization phenomena.
\newblock {\em Rev. Mod. Phys.}, 77:137--185, 2005.

\bibitem{sync1}
F.~A. Rodrigues, T.~K.~DM. Peron, P.~Ji, and J.~Kurths.
\newblock The kuramoto model: in complex networks.
\newblock {\em Phys. Rep.}, 610:1--98, 2016.

\bibitem{sync2}
S.-Y. Ha, D.~Ko, J.~Park, and X.~Zhang.
\newblock Collective synchronization of classical and quantum oscillators.
\newblock {\em EMS Surv. Math. Sci.}, 3:209--267, 2016.

\bibitem{sync3}
N.~George, F.~Bullo, and O.~Camp\`{a}s.
\newblock Connecting individual to collective cell migration.
\newblock {\em Sci. Rep.}, 7:9720, 2017.

\bibitem{sync4}
N.~Uchinda.
\newblock Many-body theory of synchronization by long-range interactions.
\newblock {\em Phys. Rev. Lett.}, 106:064101, 2011.

\bibitem{sync5}
G.~Manzano, F.~Galve, G.~L. Giorgi, E.~Hern\'{a}ndez-Garcia, and R.~Zambrini.
\newblock Synchronization, quantum correlations and entanglement in oscillator
  networks.
\newblock {\em Sci Rep.}, 3:1439, 2013.

\bibitem{sync6}
K.~P. O'Keeffe, H.~Hong, and S.~H. Strogatz.
\newblock Oscillators that sync and swarm.
\newblock {\em Nature Commun.}, 8:1504, 2017.

\bibitem{sync7}
Florian Dörflera, Michael Chertkovb, and Francesco Bulloa.
\newblock Synchronization in complex oscillator networksand smart grids.
\newblock {\em Proc. Natl. Acad. Sci. USA}, 110:2005--2010, 2013.

\bibitem{sync8}
A.~Camilli and B.~L. Bassler.
\newblock Bacterial small-moleculesignaling pathways.
\newblock {\em Science}, 311:113--116, 2006.

\bibitem{sync9}
Shir Shahal, Ateret Wurzberg, Inbar Sibony, Hamootal Duadi, Elad Shniderman,
  Daniel Weymouth, Nir Davidson, and Moti Fridman.
\newblock Synchronization of complex human networks.
\newblock {\em Nature Commun.}, 11:3854, 2020.

\bibitem{memory}
J.~Fell and N.~Axmacher.
\newblock The role of phase synchronization in memory processes.
\newblock {\em Nature Rev. Neuroscience}, 12:105--118, 2011.

\bibitem{Scholes2020}
G.~D. Scholes.
\newblock Polaritons and excitons: Hamiltonian design for enhanced coherence.
\newblock {\em Proc. R. Soc. A}, 476:20200278, 2020.

\bibitem{Ballentine}
L.~E. Ballentine.
\newblock {\em Quantum Mechanics: A modern development}.
\newblock World Scientific, Singapore, 2015.

\bibitem{Peres}
A.~Peres.
\newblock {\em Quantum Theory: Concepts and Methods}.
\newblock Kluwer, Dordrecht, 1995.

\bibitem{Barnett}
Stephen~M. Barnett.
\newblock {\em Quantum Information}.
\newblock Oxford University Press, Oxford, 2009.

\bibitem{Horodecki2009}
Ryszard Horodecki, Pawel Horodecki, Michal Horodecki, and Karol Horodecki.
\newblock Quantum entanglement.
\newblock {\em Rev. Mod. Phys.}, 81(2):865--942, 2009.

\bibitem{Vedral2002}
V~Vedral.
\newblock The role of relative entropy in quantum information theory.
\newblock {\em Rev. Mod. Phys.}, 74(1):197--234, 2002.

\bibitem{WeijunQIS1}
W~Wu and GD~Scholes.
\newblock Foundations of quantum information for physical chemistry.
\newblock {\em J. Phys. Chem. Lett.}, xx:https://arxiv.org/abs/2311.12238,
  2024.

\bibitem{Diestel}
R.~Diestel.
\newblock {\em Graph Theory}.
\newblock Springer, Hamburg, 2017.

\bibitem{Janson2000}
S.~Janson, T.~{\L}uczak, and A.~Ruci{\'n}ski.
\newblock {\em Random Graphs}.
\newblock Wiley Interscience, New York, 2000.

\bibitem{Bollobas2001}
B.~Bollob{\'a}s.
\newblock {\em Random Graphs}.
\newblock Cambridge University Press, Cambridge, 2001.

\bibitem{Lodi2021}
M.~Lodi, F.~Sorrentino, and M.~Storace.
\newblock One-way dependent clusters and stability of cluster synchronization
  in directed networks.
\newblock {\em Nature Commun.}, 12:4073, 2021.

\bibitem{graphevals}
Zoran Stani\'c.
\newblock {\em Inequalities for Graph Eigenvalues (London Mathematical Society
  Lecture Note Series, Series Number 423)}.
\newblock Cambridge University Press, Cambridge, 2015.

\bibitem{Sarnak2004}
Peter Sarnak.
\newblock What is an expander?
\newblock {\em Notices AMS}, 57:762--763, 2004.

\bibitem{expandersguide}
M.~Krebbs and A.~Shaheen.
\newblock {\em Expander Families and Cayley Graphs: A beginner’s guide}.
\newblock Oxford University Press, Oxford, 2011.

\bibitem{Lubotzky}
A.~Lubotsky.
\newblock Expander graphs in pure and applied mathematics.
\newblock {\em Bull. Amer. Math. Soc.}, 49:113--162, 2012.

\bibitem{Expanders}
S.~Hoory, N.~Linial, and A.~Wigderson.
\newblock Expander graphs and their applications.
\newblock {\em Bull. Amer. Math. Soc.}, 43:439--561, 2006.

\bibitem{Expanders2}
N.~Alon.
\newblock Explicity expanders of every degree and size.
\newblock {\em Combinatorica}, 41:447--463, 2021.

\bibitem{Alon1986}
N.~Alon.
\newblock Eigenvalues and expanders.
\newblock {\em Combinatorica}, 6:83--96, 1986.

\bibitem{Nilli1991}
A.~Nilli.
\newblock On the second eigenvalue of a graph.
\newblock {\em Discr. Math.}, 91:207--210, 1991.

\bibitem{KazhdanTbook}
Bachir Bekka, Pierre de~la Harpe, and Alain Valette.
\newblock {\em Kazhdan's Property (T)}.
\newblock Cambridge University Press, 2008.

\bibitem{Townsend2023}
Pedro Abdalla, Afonso~S. Bandeira, Martin Kassabov, Victor Souza, Steven~H.
  Strogatz, and Alex Townsend.
\newblock Expander graphs are globally synchronizing.
\newblock {\em submitted}, xx:https://doi.org/10.48550/arXiv.2210.12788, 2023.

\bibitem{KadRing}
Richard~V. Kadison and John~R. Ringrose.
\newblock {\em Fundamentals of the theory of operator algebras}.
\newblock American Mathematical Society, Providence RI, 1997.

\bibitem{Zaslavsky1}
T.~Zaslavsky.
\newblock Biased graphs. i. bias, balance, and gains.
\newblock {\em J. Comb. Theory Ser. B}, 47:32--53, 1989.

\bibitem{Reff2012}
N.~Reff.
\newblock Spectral properties of complex unit gain graphs.
\newblock {\em Lin. Alg. Appl.}, 436:3165--3176, 2012.

\bibitem{Mehatari}
Ranjit Mehatari, M.~Rajesh Kannan, and Aniruddha Samanta.
\newblock On the adjacency matrix of a complex unit gain graph.
\newblock {\em Lin. Multilin. Algebra}, 70:1798--1813, 2022.

\bibitem{Khrennikov2018SciRep}
Andrei Khrennikov, Irina Basieva, Emmanuel~M. Pothos, and Ichiro Yamato.
\newblock Quantum probability in decisionmaking from quantum
  informationrepresentation of neuronal states.
\newblock {\em Sci. Rep.}, 8:16225, 2018.

\bibitem{Barabasi}
Albert-L\'aszl\'o Barab\'asi.
\newblock {\em Network Science}.
\newblock Cambridge University Press, Cambridge, 2016.

\bibitem{Mirkovic2017}
T.~Mirkovic, E.~E. Ostroumov, J.~M. Anna, R.~van Grondelle, Govindjee, and
  G.~D. Scholes.
\newblock Light absorption and energy transfer in the antenna complexes of
  photosynthetic organisms.
\newblock {\em Chem. Rev.}, 117:249--293, 2017.

\bibitem{Scholes2003}
Gregory~D. Scholes.
\newblock Long-range resonance energy transfer in molecular systems.
\newblock {\em Annu. Rev. Phys. Chem.}, 54:57--87, 2003.

\bibitem{Ozawa1985}
M.~Ozawa.
\newblock Concepts of conditional expectations in quantum theory.
\newblock {\em J. Math. Phys.}, 28:1948--1955, 1985.

\bibitem{KB2013}
A.~Khrennikov and I.~Basieva.
\newblock Quantum model for psychological measurements: From the projection
  postulate to interference of mental observables represented as positive
  operator valued measures.
\newblock {\em NeuroQuantology}, 12:324--336, 2014.

\bibitem{Ozawa2020}
M.~Ozawa and A.~Khrennikov.
\newblock Application of theory of quantum instruments to psychology:
  Combination of question order effect with response replicability effect.
\newblock {\em Entropy}, 22:37, 2020.

\bibitem{Khrennikov2016}
A.~Khrennikov.
\newblock Quantum bayesianism as the basis of general theory of
  decision-making.
\newblock {\em Phil. Trans. R. Soc. A}, 374:20150245, 2016.

\bibitem{Sarnak1988}
A.~Lubotzky, R.~Phillips, and P.~Sarnak.
\newblock Ramanujan graphs.
\newblock {\em Combinatoria}, 8:261--277, 1988.

\bibitem{Ramanujangraphs}
Giuliana Davidoff, Peter Sarnak, and Alain Valette.
\newblock {\em Elementary Number Theory, Group Theory and Ramanujan Graphs
  (London Mathematical Society Student Texts, Series Number 55)}.
\newblock Cambridge University Press, Cambridge, 2003.

\bibitem{Lubotzky2019}
A.~Lubotzky and O.~Perzanchevski.
\newblock From ramanujan graphs to ramanujan complexes.
\newblock {\em Phil. Trans. R. Soc. A}, 378:20180445, 2019.

\bibitem{networks1}
F.~R.~K. Chung.
\newblock On concentrators, superconcentrators, generalizers, and nonblocking
  networks.
\newblock {\em The Bell System Technical Journal}, 58:1765--1777, 1979.

\bibitem{networks2}
Ernesto Estrada, Naomichi Hatanoe, and Michele Benzi.
\newblock The physics of communicability in complex networks.
\newblock {\em Phys. Rep.}, 514:89--119, 2012.

\bibitem{osc-circuit}
Heng-Chia Chang, Eric~S. Shapiro, and Robert~A. York.
\newblock Influence of the oscillator equivalent circuit on the stable modes of
  parallel-coupled oscillators.
\newblock {\em IEEE Trans. Microwave Theory and Techniques}, 45:1232--1239,
  1997.

\bibitem{nanotech}
Damir Vodenicarevic, Nicolas Locatelli, Flavio~Abreu Araujo, Julie Grollier,
  and Damien Querlioz.
\newblock A nanotechnology-ready computing scheme based on aweakly coupled
  oscillator network.
\newblock {\em Sci. Rep.}, 7:44772, 2016.

\bibitem{spiking1}
Kashu Yamazaki, Viet-Khoa Vo-Ho, Darshan Bulsara, and Ngan Le.
\newblock Spiking neural networks and their applications: A review.
\newblock {\em Brain Sciences}, 12:863, 2022.

\bibitem{hairballs}
Koon-Kiu Yan, Daifeng Wang, Anurag Sethi, Paul Muir, Robert Kitchen, Chao
  Cheng, and Mark Gerstein.
\newblock Cross-disciplinary network comparison: Matchmaking between hairballs.
\newblock {\em Cell Systems}, 2:147--157, 2016.

\bibitem{brain1}
S.~N. Baker, J.M. Kilner, E.M. Pinches, and R.N. Lemon.
\newblock The role of synchrony and oscillations in the motor output.
\newblock {\em Exp. Brain Res.}, 128:109--117, 1999.

\bibitem{brain2}
J.~F. Cavanagh and M.~J. Frank.
\newblock Frontal theta as a mechanism for cognitive control.
\newblock {\em Trends Cognitive Sci.}, 18:414--421, 2014.

\bibitem{brain3}
A.-L. Giraud and D.~Poeppel.
\newblock Cortical oscillations and speech processing: emerging computational
  principles and operations.
\newblock {\em Nature Neurosci.}, 15:511--517, 2012.

\bibitem{brain4}
J.~E. Lisman and O.~Jensen.
\newblock The theta-gamma code.
\newblock {\em Neuron}, 77:1002--1016, 2013.

\bibitem{brain5}
M.~Siegel, T.H. Donner, and A.K. Engel.
\newblock Spectral fingerprints of large-scale neuronal interactions.
\newblock {\em Nature Rev. Neurosci.}, 13:121--134, 2012.

\bibitem{brain6}
P.~Fries.
\newblock Rhythms for cognition: Communication through coherence.
\newblock {\em Neuron}, 88:220--235, 2015.

\bibitem{brain7}
R.~VanRullen.
\newblock Perceptual cycles.
\newblock {\em Trends Cognitive Sci.}, 20:723--735, 2016.

\bibitem{monkey1}
Nicolas Brunet, Conrado~A. Bosman, Mark Roberts, Robert Oostenveld, Thilo
  Womelsdorf, Peter~De Weerd, and Pascal Fries.
\newblock Visual cortical gamma-band activity during free viewing of natural
  images.
\newblock {\em Cerebral Cortex}, 25:918–926, 2015.

\bibitem{corticalnetworks}
Joerg~F. Hipp, Andreas~K. Engel, and Markus Siegel.
\newblock Oscillatory synchronization in large-scalecortical networks predicts
  perception.
\newblock {\em Neuron}, 69:387–396, 2011.

\end{thebibliography}

\end{document}